\documentclass[12pt]{article}
\usepackage{euscript,graphicx,epigraph,amsmath}
\newcommand{\karti}[4]{\begin{figure}[#1]\begin{center}
\includegraphics[height=#2]{#3.ps}
\end{center}\caption{#4}\end{figure}}

\DeclareMathOperator{\Bd}{Bd}
\newcommand{\ssy}[5]{#1, \emph{#2} \textbf{#3}, #5 (#4)}
\DeclareSymbolFont{symbols}     {OMS}{ztmcm}{m}{n}
\newcommand{\R}{\ensuremath{\mathcal R}}
\newcommand{\G}{\ensuremath{\mathcal  G}}
\renewcommand{\textflush}{flushleftright}
\begin{document}
\title{Time machine (1988--2001)}
\author{S. Krasnikov}
\date{}
\maketitle

\epigraph{``...don't drive yourself crazy trying to resolve
the paradoxes of time travel."}{L. S. de Camp \cite{ep}}
\noindent The time machine (in its science fiction incarnation) was
born back in 19-th century \cite{Wells}, i.~e.\  well before  the
concept of spacetime appeared. However, it was
relativity that provided an adequate language for discussing and
studying time machines. In particular, according to special
relativity, our world is the Minkowski space and the life history of a
pointlike particle is a curve in this space, see Fig.~\ref{fig1}.
\karti{h}{0.4\textwidth}{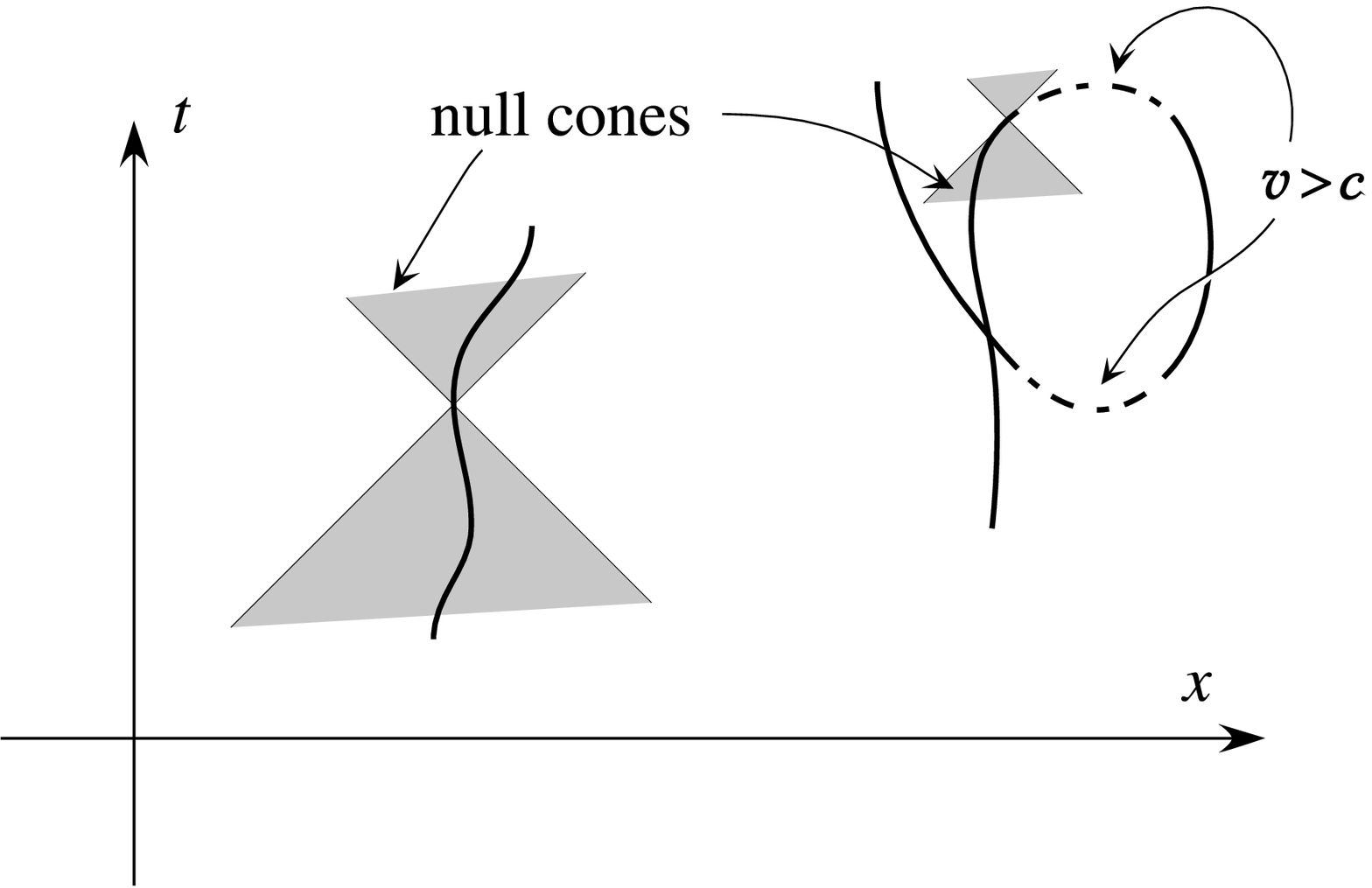}{No time machine in special
relativity.\label{fig1}} In these terms one can formulate, in quite a
meaningful way, the question about the existence of time machines, or
--- which is the same  --- about the possibility of an observer
meeting his older self. Namely, one can ask, whether a world line of a
particle may have a self-intersection. The answer is straightforward.
To make a loop a curve must somewhere leave the null cone as shown in
Fig.~\ref{fig1}. A particle with such a world line would exceed the
speed of light and, insofar as no tachyons have been found yet, the time
machine is impossible in special relativity.

In general relativity the situation is much less trivial.
\karti{hbt}{0.5\textwidth}{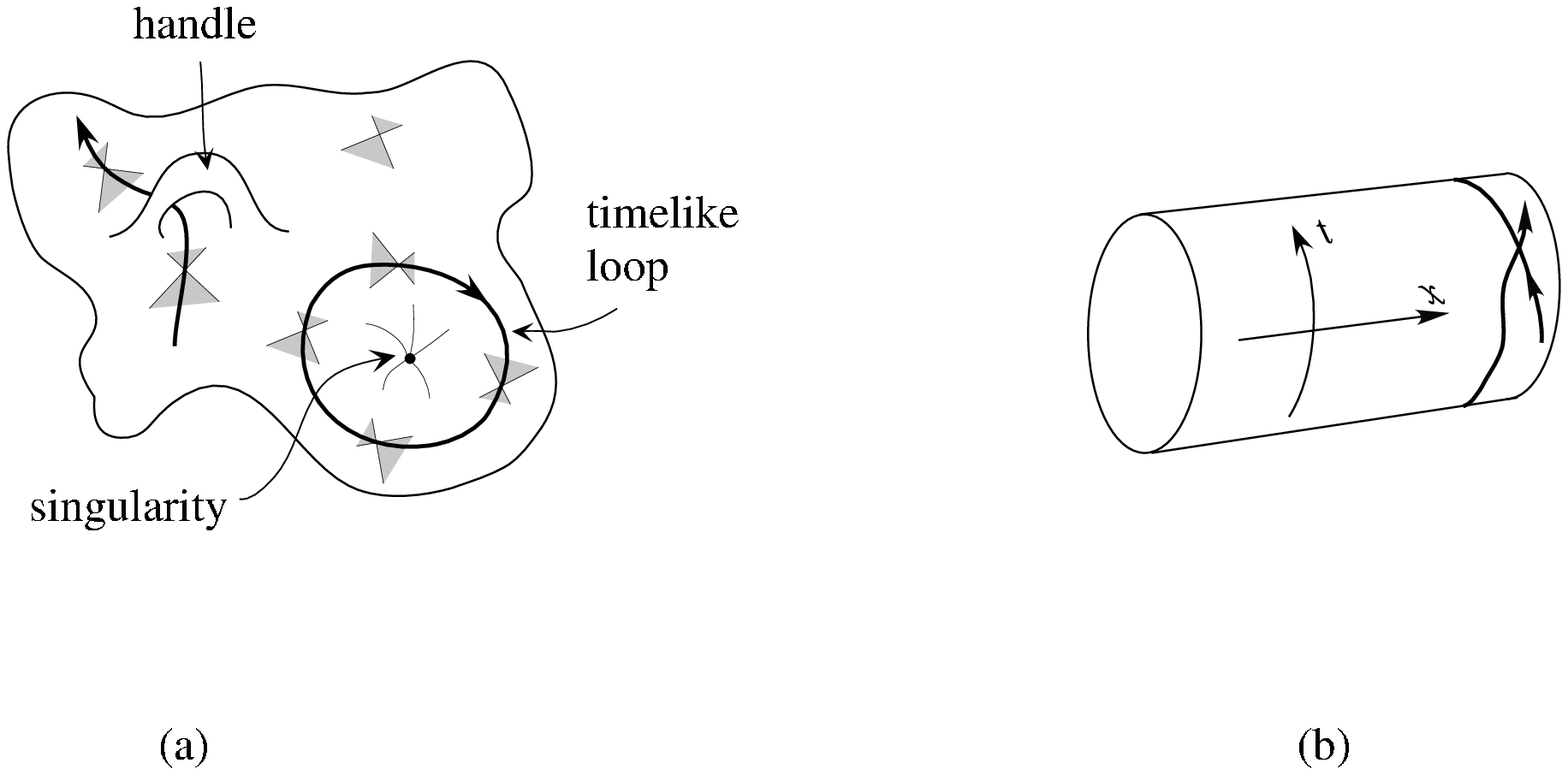}{\label{fig2} Timelike loops
due to the non-trivial geometry (\emph{a}), or topology (\emph{b}) of
the spacetime.} According to this theory, our spacetime must be a smooth
Lorentzian manifold, i.~e.\  in sufficiently small regions it must be
`approximately Minkowskian', but at large scale (as long as the
Einstein equations with a reasonable matter source hold), it may have
any geometry and topology. There may be holes, handles, see
Fig.~\ref{fig2}a; almost whatever one wants. And the null cones need
not any more look all in the same direction. In particular, spacetimes
are conceivable with closed or self-intersecting timelike curves.

A simple example is the Minkowski space rolled into a cylinder, see
Fig.~\ref{fig2}b. Locally everything is fine in this spacetime, but
due to its non-trivial global structure, an observer can meet his
younger self. Such spacetimes \cite{Goedel,Tipler}, however, do not
deserve the name `time machine', because the closed timelike curves
\emph{always} exist in them. They are not \emph{created} at some moment.
In this sence, a better example is the Deutsch--Politzer (DP) space
\cite{Deu}, which we shall describe as a result of the following
surgery. Make two cuts in the Minkowski space
\karti{hbt}{0.42\textwidth}{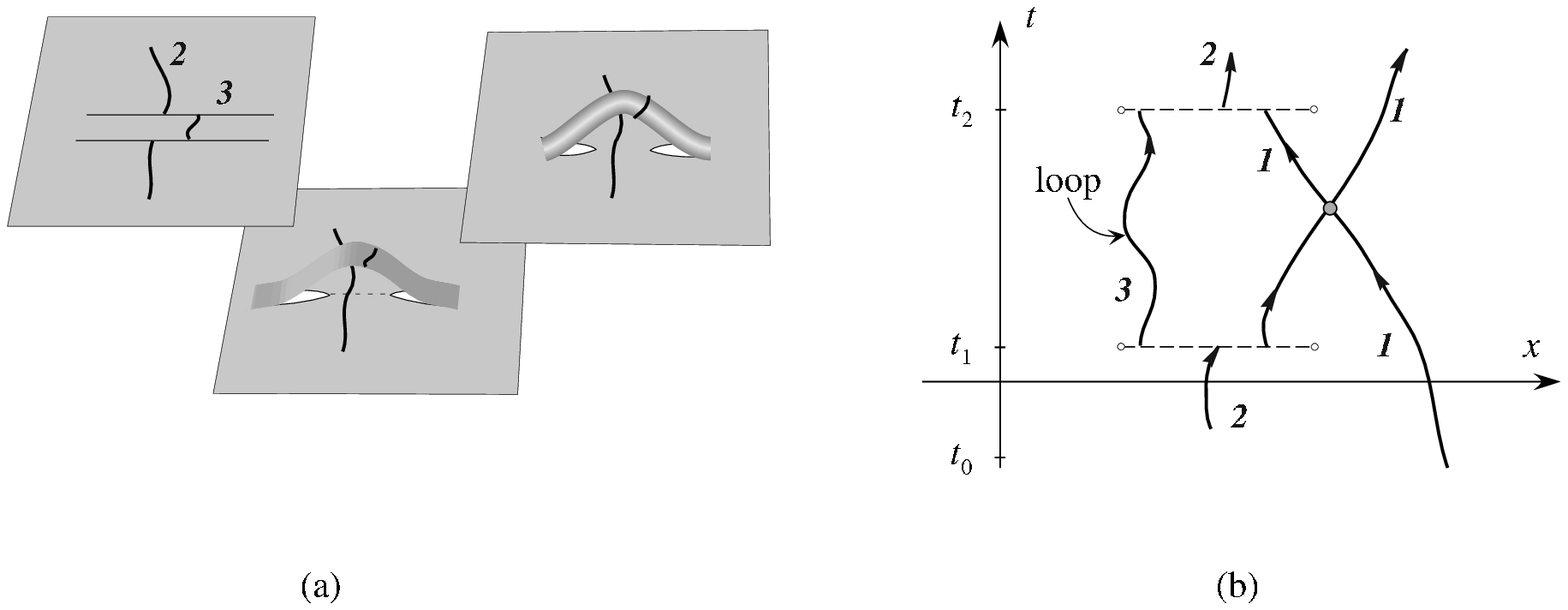}{\label{fig:dp}(\emph{a}) Preparation of
the DP spacetime from the Minkowski plane. (\emph{b}) All lines
(1,2,3) are actually continuous.}
and glue the upper bank of lower cut
to the lower bank of the upper cut and vice verse, as shown in
Fig.~\ref{fig:dp}a, so that  a cylinder appears attached to the plane.
It is convenient to  draw the resulting spacetime  still as the
Minkowski plane, see Fig.~\ref{fig:dp}b, and just to keep in mind that
any curve reaching the lower segment must be continued from the upper
and, vice verse.

The DP space contains timelike loops, which all lie to the future  of
a good (causal) region, so it fits the name `time machine'. This
model, of course, in unsuitable for tackling the problem of
\emph{building} a time machine (it gives no clue into what could
\emph{force} the spacetime to evolve into the time machine instead of
just remaining Minkowskian, see below), but is quite adequate in
studying what can happen, if for some reason or other a time machine
did appear.
\par
One can often read that the existence of a time machine would lead to
awful paradoxes  contradicting  common sense and the notion of free
will, or even proving the impossibility of the time machine. Now  we
can verify this assertion. Let us begin with the notorious
grandfather paradox, which is a story of a researcher \R, who asks
his grandson
\G\ to trip (by time machine) to the past and to kill him (\R) in
infancy. The grandson tries to grant his wish
\karti{hbt}{0.54\textwidth}{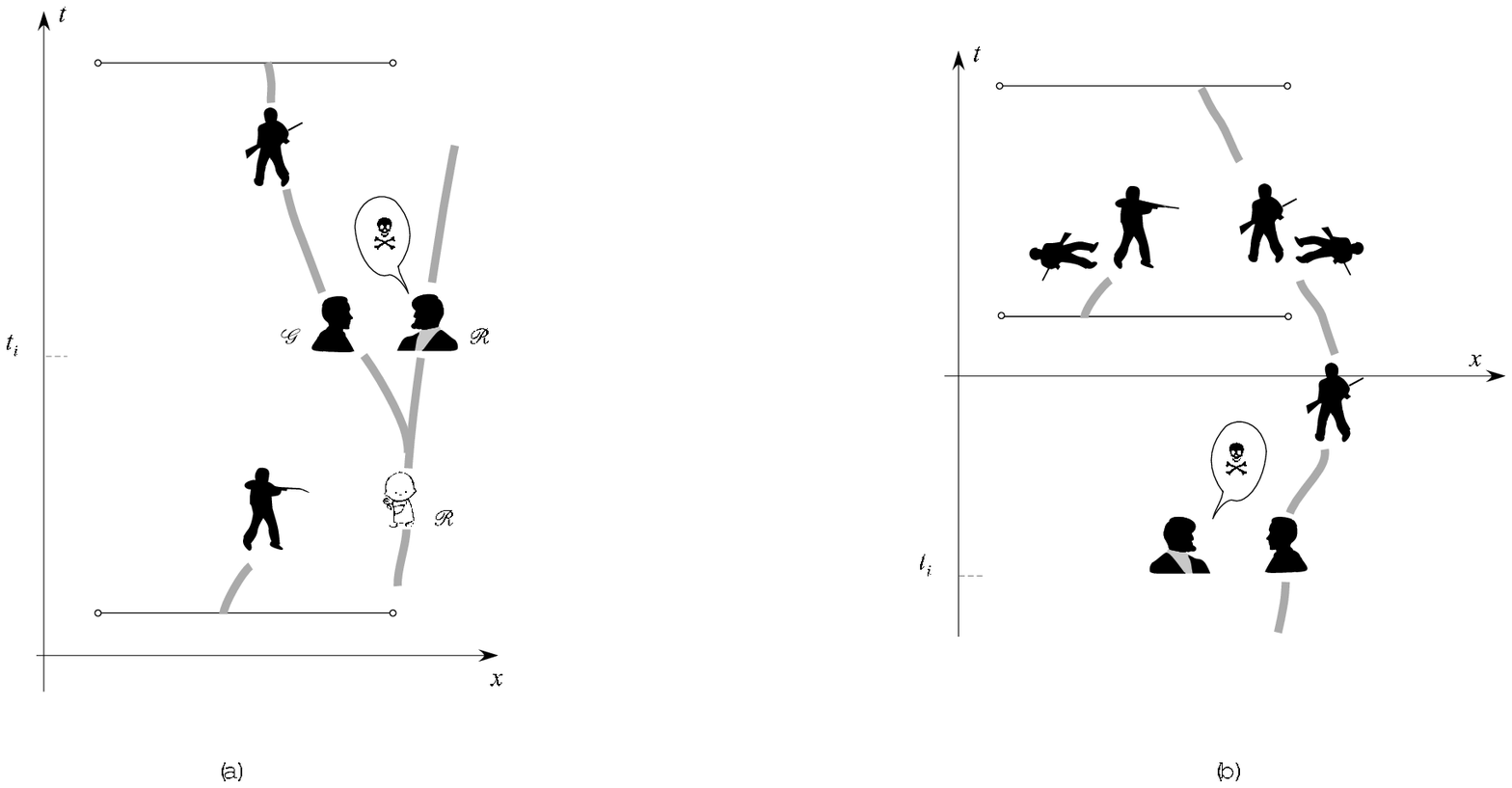}{\label{fig3} A paradox admits a
\emph{simple} solution, when the initial conditions being fixed at
$t_i\in(t_1,t_2)$ are affected by the outcome of the experiment
(\emph{a}), and \emph{none} otherwise (\emph{b}).} (see
Fig.~\ref{fig3}a), but the problem is: if
\G\ succeeds than the baby does not  grow up and, correspondingly,
does not have any grandchildren that could kill him. Which proves that
the grandson \emph{fails} to kill the baby. The circumstances are not
detailed, so there is much room for explanations \emph{just why} he
fails.

Say, the gun can jam, or the shooter can miss. However, if we arm
grandsons again and again (hundreds of times) and
\emph{always} only bad shooters get  reliable  guns, this, of course, would
look miraculous. Why cannot we give a good gun to a marksman?
Is not this a restriction on our freedom of will?

In fact, the solution of the grandfather paradox is quite simple. The
point is that  only those persons are \emph{available} for our
experiment whose grandfathers by whatever reasons were \emph{not}
killed. So,  what prevents us from arming a sufficiently good shooter
with a sufficiently good gun is not problems with our freedom of
will\footnote{The grandson, being born with already known future life,
is such an unusual object that we need not worry about \emph{his}
freedom of will.}, but just the shortage of such shooters. And the
better our guns are the fewer  marksmans are at our disposal.

A more detailed analysis shows \cite{ttrpar} that to avoid such simple
solutions one must consider experiments performed \emph{before} the
time machine appeared \cite{Deu}. Indeed, suppose one kindly requests
a person to wait until a time machine is built (at some time in the
\emph{future}), to enter it, and to  kill his (the traveler's) younger
self (see Fig.~\ref{fig3}b).
The situation in the time machine is exactly as paradoxical as in the
grandfather case --- if the test person is dead when he meets his
younger self, he cannot kill the latter, so why he is dead? And if he
is alive and shooting, then how come the victim survived?
This time, however, the initial conditions do not depend on what
occurs in the time machine and the paradox  has no simple resolutions.
Moreover, being  formulated in terms of pointlike particles (which
allows one to check \emph{all} possibilities) it has no resolutions
\emph{at all} \cite{ttrpar}.

How to fight the time travel paradox? The most obvious way out would
be to forbid the time machines by fiat, to \emph{postulate} causality.
But this is not so easy. In 1988 Morris, Thorne, and Yurtsever (MTY)
published a paper, where they considered a spacetime $M$ shown in
Fig.~\ref{fig:MTY}.
\karti{hbt}{0.52\textwidth}{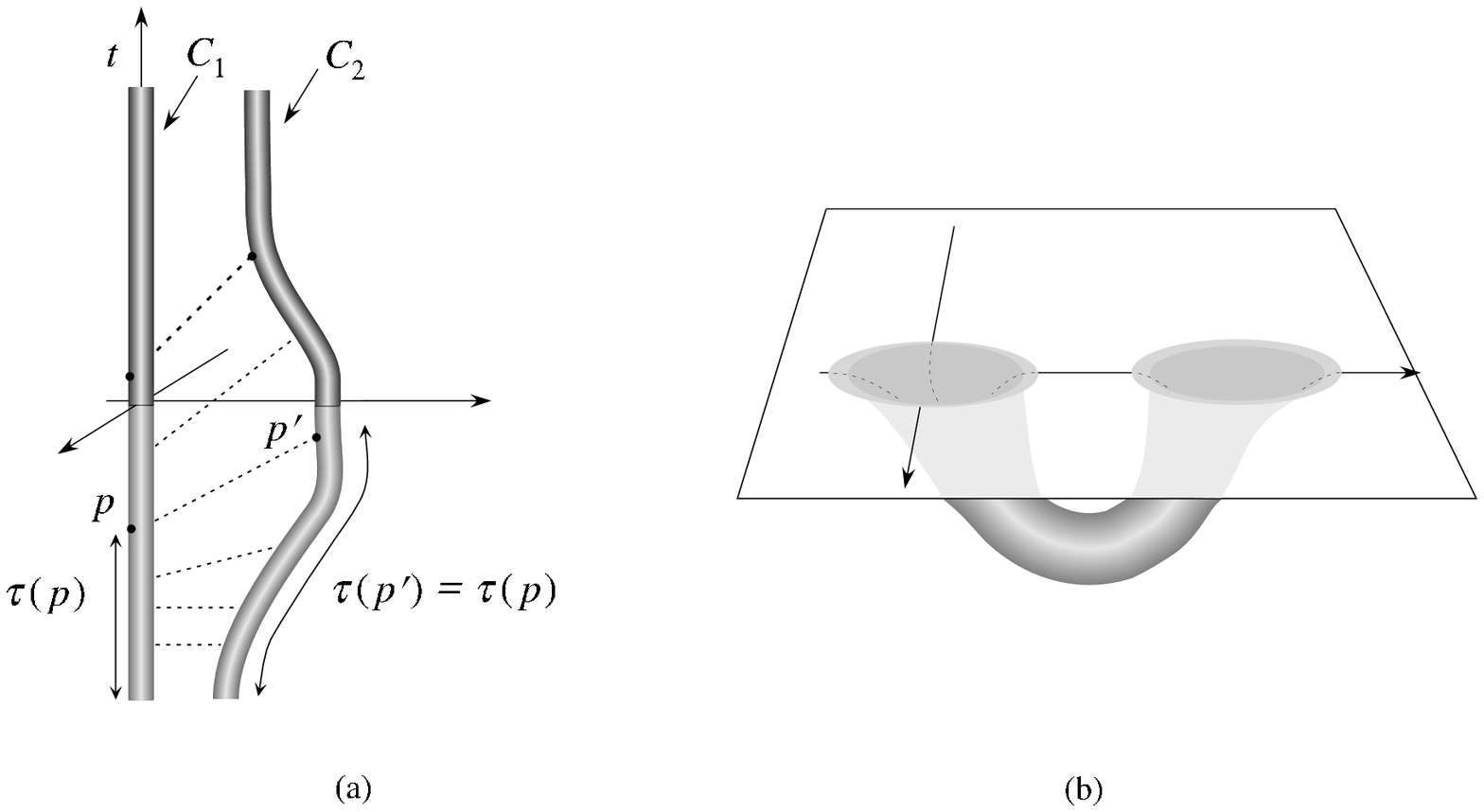}{\label{fig:MTY}(\emph{a}) The dashed
lines connect the identified points. (\emph{b}) The section $t=-2$ of
$M$.} Geometrically $M$ can be obtained by removing  two close
cylinders $C_1$ and $C_2$ from the Minkowski space and gluing together
the boundaries $\EuScript B_{1,2}\equiv\Bd C_{1,2}$  of the resulting
holes (the junction is then smoothed out by curving appropriately a
close vicinity of $\EuScript B_1=\EuScript B_2$). The cylinders are
taken to be parallel to the $t$-axis except that $C_2$ has a bend at,
say, $-1<t<1$. The gluing must respect the following rule
\begin{equation}\label{eq:rule}\tag{$*$}
p=p'\quad\Rightarrow\quad \tau(p)=\tau(p'),
\end{equation}
where $\tau(x)$ is defined to be the length of the longest (recall
that the metric is Lorentzian) timelike curve lying in $\EuScript
B_{1,2}$ and connecting $x$ with the surface $t=-2$. Due to the bend,
$\tau$ grows with time slower on $\EuScript B_{2}$ than on $\EuScript
B_{1}$ (the `twin paradox'). Correspondingly, some of the identified
points were causally related in the `initial' (Minkowski) space. So,
there are timelike loops in $M$. Physically, $M$ presents a spacetime
in which a wormhole (see Fig.~\ref{fig:MTY}b) evolves so that the
distance (in the ambient flat space) between its mouths changes at
$-1<t<1$ without significant changes in the form or length of the
throat (it is this last condition that necessitates the rule
\eqref{eq:rule}).

The MTY time machine cannot be banished as easily as the DP one,
because we have (in assumption that wormholes exist, they are large
and stable, etc.) a specific prescription of how to build it.
Precisely this was the point of \cite{MTY}:  assuming the time
machines \emph{are} prohibited,  exactly how this prohibition is
enforced? Suppose, one finds a wormhole, pushes one of its mouths,
then pulles it back. What \emph{exactly} will go wrong? What will
prevent a closed causal curve from appearance? It was such statement
of the problem that gave birth to the time machine as an element of
the physical, rather than philosophical, or science fiction realm.

A mechanism that could prevent the appearance  of a closed timelike
curve was sought for more than a decade, but has never been found
\cite{rev}. In this sense causality still remains unprotected
\cite{Conj}. But does it mean that an advanced civilization can build a
time machine and face the ensuing paradoxes? No, it does not. The
point is that there is a great difference between \emph{observing} a
time machine appearing by itself and \emph{creating} one. And while
the former seemingly is possible, the latter is not. The difference is
well exemplified by the DP space. We can prepare a flat region in our
spacetime and, for all we know, a time machine well may appear there.
But just as well it may not, and we shall have a usual Minkowski
space. The choice is up to the spacetime and we can do nothing to
affect it. Surprisingly, exactly the same is true for \emph{any} time
machine, be it based on a wormhole \cite{MTY}, a cosmic string
\cite{Gott}, or whatever else \cite{Conj}.  A theorem \cite{eotm}
proved in 2001
says:
\emph{Any  spacetime obeying local laws and  having no causal loops
in its past, has a causal maximal extension.} Which means (since the
Einstein equations are local) that within general relativity a time
machine cannot be built.

\end{document}